\documentclass[amsmath,superscriptaddress,prl,twocolumn] {revtex4}
\usepackage{bm}
\usepackage{enumerate}

\usepackage{graphicx}
\usepackage[dvips]{epsfig}
\usepackage{epsf}

\makeatletter
\newcommand{\Rmnum}[1]{\expandafter\@slowromancap\romannumeral #1@}
\makeatletter

\begin{document}
\title{Metallic collinear antiferromagnets with mirror-symmetric and asymmetric spin-splittings}
\author{Vladimir A. Zyuzin}
\affiliation{L.D. Landau Institute for Theoretical Physics, 142432, Chernogolovka, Russia}
\begin{abstract}
In this paper we theoretically describe a distinct class of two-dimensional N\'{e}el ordered metallic antiferromagnets on a honeycomb-like lattice in which the two sublattices are connected only by a combination of time-reversal and mirror symmetry operations.
As a result of this symmetry, conducting fermions have antiferromagnetic spin-splitting consistent with the symmetry, the mirror-symmetric spin-splitting. It is shown that the anomalous spin Hall effect is expected in such systems.
We also consider a system in which there are no symmetries between the sublattices and obtain asymmetric spin-splitting. Such systems are expected to have the anomalous Hall effect. 
\end{abstract}
\maketitle

It has been understood that local non-magnetic environment of the two sublattices in N\'{e}el ordered metallic antiferromagnets might result in $d-$, $g-$, or $i-$ wave symmetric spin-splitting of conducting fermions \cite{HayamiYanagiKusunose2019,AHE_AFM,Rashba2020,HayamiYanagiKusunose2020,SmejkalSinovaJungwirth2022a,SmejkalSinovaJungwirth2022b,AHE_AFM_Review}. Such spin-splitting occurs because the two sublattices forming the N\'{e}el order are connected only via time-reversal and rotation operations.
It has been suggested to call such spin-splitting of conducting fermions as the altermagnetism of N\'{e}el ordered metallic antiferromagnets \cite{SmejkalSinovaJungwirth2022a,SmejkalSinovaJungwirth2022b}. 
The spin-splitting of conducting fermoins is antiferromagnetic, and there is no net magnetization. 
However, finite magnetic moment can appear upon addition of spin-orbit coupling to the altermagnetic structure. 

Magnetic moment in antiferromagnetic altermagnets can be understood as the Dzyaloshinskii's weak ferromagnetism \cite{Dzyaloshinskii1958}, which appears in N\'{e}el ordered insulating antiferromagnets due to the spin-coupling canting the spins in the N\'{e}el order, but with the only difference that it is now the conducting fermions which are responsible for the magnetic moment due to the spin-orbit coupling while the N\'{e}el order is intact.

In Ref. \cite{HayamiYanagiKusunose2020,Brekke2023,AgterbergPRB2024} different minimial models of altermagnetic coupling have been proposed.
In addition, in Ref. \cite{Zyuzin2024} the author has proposed a simple microscopic model of $\textit{d}$-wave symmetric altermagnetic coupling of conducting fermions in antiferromagnet with the N\'{e}el order on a two-dimensional square lattice of the checkerboard type. The $\textit{d}$-wave symmetric spin-splitting was shown to be originating due to the interplay of interaction of conducting fermions with the N\'{e}el order and their anisotropic second-nearest neighbor hopping. What was essential in the mechanism is the anisotropic second-nearest neighbor hopping achieved by the non-magnetic atoms placed on the square lattice in the checkerboard pattern. Such non-magnetic atoms reduce fermion hopping across them and might even block it. 
We point out that the altermagnetism of conducting fermions in the scenario of Ref. \cite{Zyuzin2024} is a byproduct of the antiferromagnetic instability of the system.

Here we wish to exploit this idea further and study N\'{e}el ordered antiferromagnet on a honeycomb lattice with patterns of non-magnetic atoms which block tunneling of conducting fermions across them. 
We deduce a distinct class of N\'{e}el ordered metallic antiferromagnets in which the two sublattices are connected only by a combination of time-reversal and mirror operations. 
As a result of the symmetry there is an antiferromagnetic spin-splitting of the conducting fermions. 
The spin-splitting is not of the $d-$, $g-$, or $i-$ wave type but rather of a distinct type which is consistent with the symmetry of the lattice, we call it as the mirror-symmetric spin-splitting. 
Obtained mirror-symmetric spin-splitting of conducting fermions is antiferromagnetic and in that respect mirror-symmetric magnetic systems are antiferromagnets. 
We suggest that breaking of the aforementioned symmetry of the magnetic system will generate the anomalous Hall effect.  
It is possible that such mirror-symmetric spin-splitting has already been experimentally observed in MnTe antiferromagnet \cite{MnTe2023exp,OrlovaDeviatov2024,Belashchenko2024}.

\begin{figure}[t] 
\includegraphics[width=0.8 \columnwidth ]{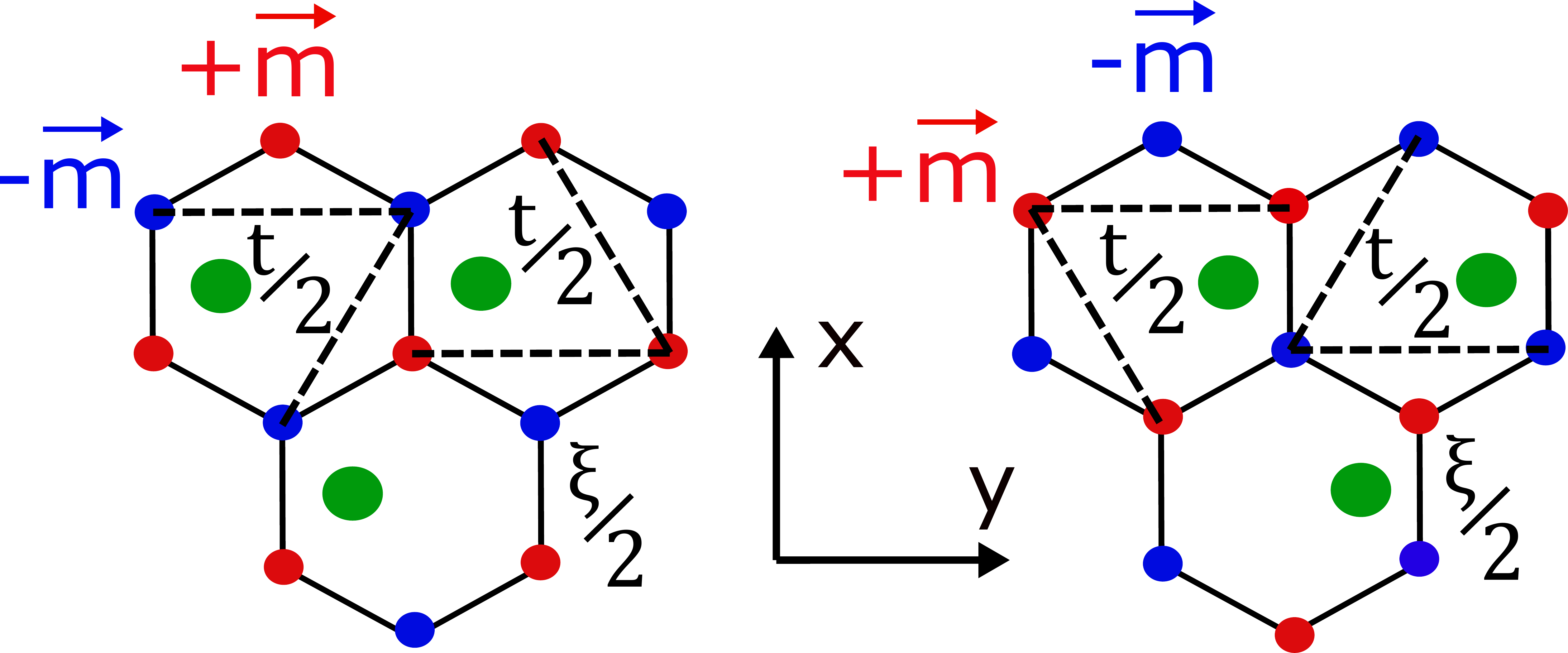} ~~~~

\protect\caption{Unit cell in the two models consists of two magnetic sites shown in red and blue and one non-magnetic shown in green. Red is for $+{\bf m}$, blue is for $-{\bf m}$. 
Green atoms are non-magnetic and we assume there is no energy state for the conduction fermions on them. 
In addition, we assume that green atoms suppress tunneling of conducting fermions across them.
Dashed lines are the only allowed second-nearest neighbor hopping of fermions.
The two (on the left and on the right) lattices are related by the mirror and time-reversal operations. 
A combination of mirror and time-reversal operations is the symmetry of the system. 
 }
\label{fig:fig1}  
\end{figure}

In order to see how this mirror-symmetric spin-splitting occurs,
let us study an antiferromagnet on the honeycomb lattice shown in Fig. (\ref{fig:fig1}). 
We assume that the N\'{e}el order is given and we treat it as non-fluctuating, that there are conducting fermions in the system and they are not affecting the N\'{e}el order and only interact with it through site-dependent exchange interaction.
Red and blue atoms are magnetic and correspond to spin-up ($+{\bf m}$) and spin-down ($-{\bf m}$) localized magnetic moments. Fermions have an energy state only on the red and blue sites. Fermions are of the $s-$wave type.
The green atom is non-magnetic, and we assume that fermion state on the green atom doesn't overlap with those on the red and blue sites, which means that fermions don't hop from red and blue sites to the green ones. In addition to this, the greem atoms are assumed to block tunneling of fermions through them. In general they reduce the tunneling through them, but for the sake of simplicity we chose the limiting case. Our results are expected to remain in the general case.
The physical space of fermions is the two sublattices and two spins, such that the overall spinor structure is $\Psi^{\dag} = (\psi^{\dag}_{{\mathrm{R},\uparrow}},\psi^{\dag}_{{\mathrm{R},\downarrow}},\psi^{\dag}_{{\mathrm{B},\uparrow}},\psi^{\dag}_{{\mathrm{B},\downarrow}})$, where $\mathrm{R}$ and $\mathrm{B}$ stand for red and blue correspondingly and arrows stand for spin. 

The Hamiltonian reads as
\begin{align}\label{ModelA}
\hat{H}_{\mathrm{A}}= &
\left[
\begin{array}{cc} 
{\bf m}\cdot{\bm \sigma}+\chi_{\mathrm{R};{\bf k}}\sigma_{z} + t_{\mathrm{R};{\bf k}}  & \xi_{\bf k} - i\eta_{\bf k} \sigma_{z}\\
\xi_{\bf k}^{*} + i\eta_{\bf k}^{*} \sigma_{z} & -{\bf m}\cdot{\bm \sigma}  + \chi_{\mathrm{B};{\bf k}}\sigma_{z}+ t_{\mathrm{B};{\bf k}}
\end{array}
\right],
\end{align}
where $\xi_{\bf k} = \xi \left[ 2e^{i\frac{k_{x}}{2\sqrt{3}}}\cos\left( \frac{k_{y}}{2} \right) + e^{-i\frac{k_{x}}{\sqrt{3}}} \right]$ is the first-nearest neighbor hopping.
Second-nearest neighbor hopping, keeping in mind that the green atom doesn't allow for hopping of fermions across it, reads as
\begin{align}
t_{\mathrm{R}/\mathrm{B};{\bf k}} 
&= t\cos\left( k_{y} \right) + t\cos\left( \frac{\sqrt{3}k_{x}}{2} \right)\cos\left( \frac{k_{y}}{2} \right) 
\nonumber
\\
&
\mp
t\sin\left( \frac{\sqrt{3}k_{x}}{2} \right)\sin\left( \frac{k_{y}}{2} \right).
\end{align}
We observe that the hopping is different for spin up and spin down fermions. 
This is the origin of the mirror-symmetric spin-splitting.
For the sake of generality we list possible spin-orbit coupling intrinsic to the lattice.
Lattice structure allows for the first-nearest neighbor spin-orbit coupling 
$\eta_{\bf k} = e^{-i\frac{k_{x}}{\sqrt{3}}}$. Second nearest neighbor spin-orbit coupling reads as 
$
\chi_{\mathrm{R}/\mathrm{B};{\bf k}} = \left( \lambda_{{\bf k}} \pm \gamma_{{\bf k}}\right), 
$
where its isotropic part  is 
\begin{align}
&
\lambda_{\bf k} = -\lambda \sin\left( \frac{\sqrt{3}k_{x}}{2} \right)\cos\left( \frac{k_{y}}{2} \right),
\end{align}
and its anisotropic part is
\begin{align}
\gamma_{\bf k} = \lambda \left[ \cos\left( \frac{\sqrt{3}k_{x}}{2} \right)\sin\left( \frac{k_{y}}{2} \right) - \sin\left( k_{y} \right) \right].
\end{align}
The anisotropic part of the spin-orbit coupling breaks accidental mirror symmetry about the $zx$ plane shown in Fig (\ref{fig:fig1}). 
If the charge of the green atom is the same as the one on red and blue, then the second-nearest spin-orbit coupling is mostly suppressed and we can set $\lambda \approx 0$. 
On the other hand, if the charge of the green atom is opposite with the one on the red and blue, the second-nearest neighbor spin-orbit coupling will be enhanced. This can be considered as a model of ferroelectric antiferromagnet.
\begin{figure}[t] 
\includegraphics[width=0.4 \columnwidth ]{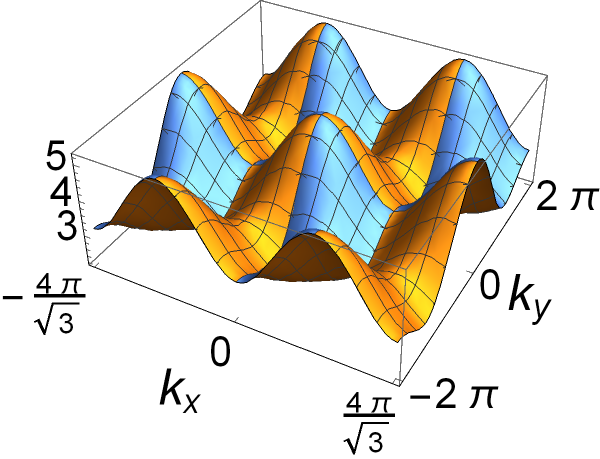} ~~~~
\includegraphics[width=0.4 \columnwidth ]{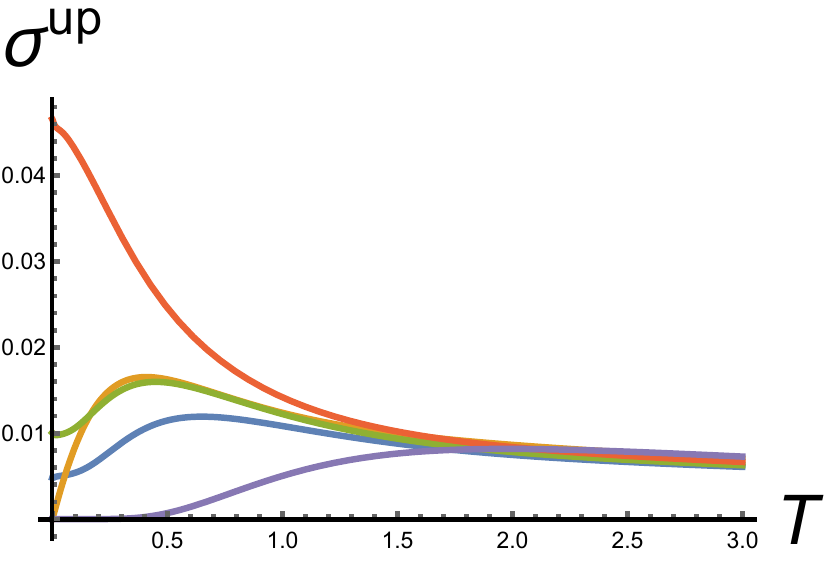}

\protect\caption{Left: spectrum of the conduction band in three times the first Brillouin zone (three pairs of ${\bf K}$ and ${\bf K}^\prime$ points are shown) for illustration purposes. Blue color corresponds to the spin-up fermions, yellow to the spin-down. Parameters are $\xi = 1$, $t=0.5$, $m=3$, $\lambda=\eta = 0$.  
Right: plot of the $\sigma^{\uparrow}$ defined after Eq. (\ref{currentAHE}) as a function of temperature for 
$\xi = 1$, $t=0.5$, $m=3$, $\eta = 0.4$, $\lambda = 0.2$ as a function of temperature $T$ for different values of $\mu$.
Blue $\mu =4$, green $\mu=3.7$, red $\mu=3$, yellow $\mu=2$, purple $\mu=0$.
All the parameters are in units of $\xi$ such that $\xi = 1$
 }
\label{fig:fig2}  
\end{figure}

\begin{figure}[t] 
\includegraphics[width=0.3 \columnwidth ]{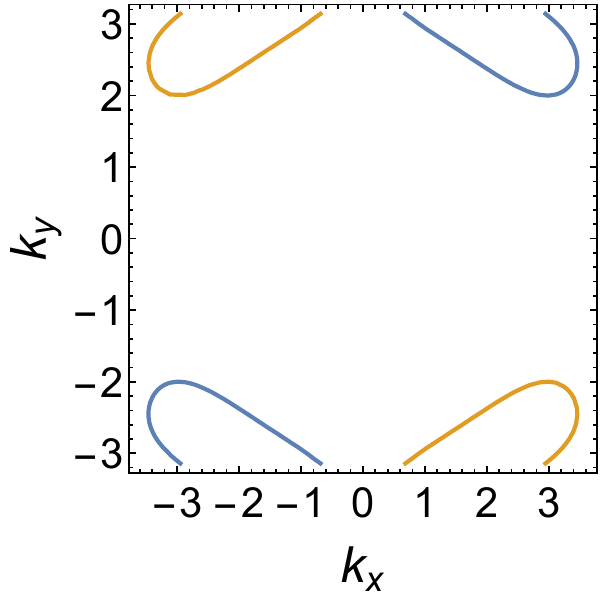} 
\includegraphics[width=0.3 \columnwidth ]{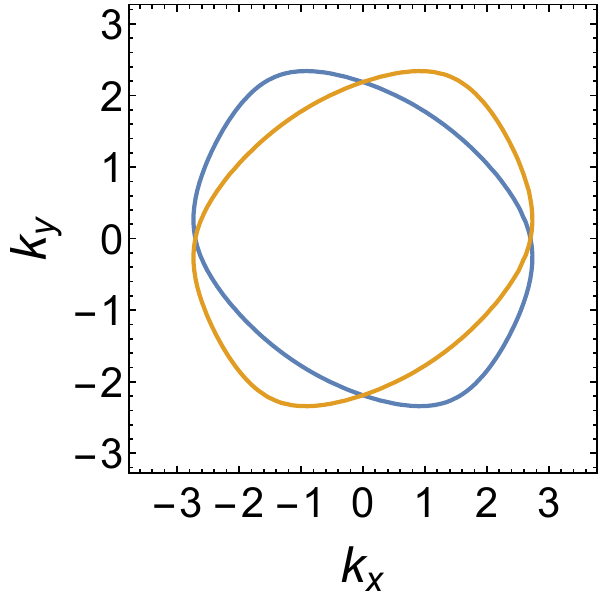} 
\includegraphics[width=0.3 \columnwidth ]{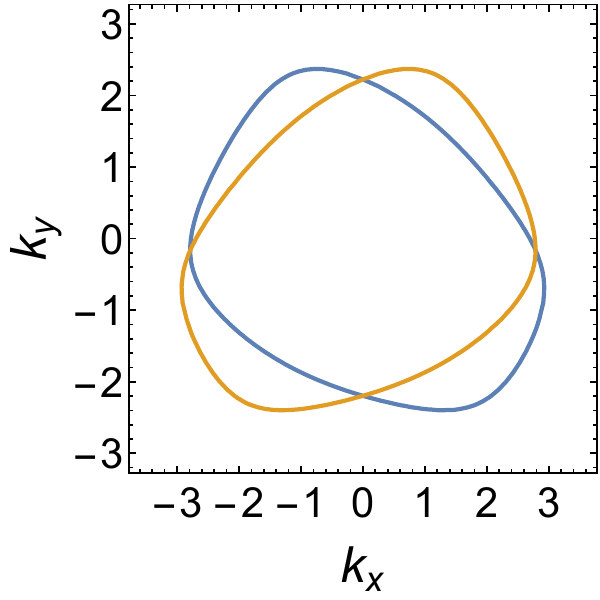} 

\protect\caption{Contour plots of the Fermi surface for chosen Fermi level $\mu$. 
There is no time-reversal and $\frac{\pi}{2}$ rotation symmetry. Blue color corresponds to the spin-up fermions, yellow to the spin-down. 
Left: in the vicinity of the ${\bf K}$ and ${\bf K}^\prime$ points for $\mu =2.4$.
Center: in the vicinity of the $\Gamma$ point for $\mu=3.5$.
Left and center are plotted for $\xi = 1$, $t=0.5$, $m=3$, $\lambda=\eta = 0$.
Right: direction-dependent spin-orbit coupling reduces the symmetry to a combination of a mirror operation in $zy$ plane and time-reversal operation. The contour is plotted for $\mu=3.5$ and $\xi = 1$, $t=0.5$, $m=3$, $\lambda = 0.2$, and $\eta=0.4$.
 }
\label{fig:fig3}  
\end{figure}
Both spin-orbit coupling terms involve only the $\sigma_{z}$ component because gradients of the uncompensated atomic potentials and fermion hopping occurs in the $xy$ plane.
We set the N\'{e}el order to be in the $z-$direction and plot the spectrum of conducting fermions and their contour at the Fermi level in Fig. (\ref{fig:fig2}) and equal energy contour plots in Fig. (\ref{fig:fig3}). The left and center plots in Fig. (\ref{fig:fig3}) are plotted for zero spin-orbit coupling. We observe that combinations of time-reversal operation together with the mirror not only in $zx$ plane but also in $zy$ plane are the symmetry of the lattice for zero spin-orbit coupling. A symmetry under the combination of time-reversal and mirror in $zx$ plane operation is accidental. Indeed, a lattice on the right in Fig. (\ref{fig:fig1}), obtained from the left one by the time-reversal and mirror in $zx$ plane operations, has the same second-nearest neighbor hoppings for spin-up and spin-down fermions as the one on the left. Addition of spin-orbit coupling breaks this accidental symmetry as can be see in the right plot of Fig. (\ref{fig:fig3}). 
Also none of the spin-orbit coupling breaks the symmetry of time-reversal and mirror in the $zy$ plane.

It is expected that the model will show the analog of the $d-$wave Hall effect \cite{VorobevZyuzin2024}, $d-$wave linear magnetoconductivity \cite{VorobevZyuzin2024}, or of the $d-$wave anomalous Hall effect in a three dimensional generalization of the model provided certain spin-orbit coupling is present and the N\'{e}el order is set to the $xy$ plane \cite{AHE_AFM}. These effects will have four zeros just like in the $d-$wave spin-splitting case. Since the orbital magnetization has the same structure as the Hall effect, it is expected that it will have the mirror-symmetric structure with four zeros as well. We think such orbital magnetization has already been experimentally observed in MnTe antiferromagnet \cite{OrlovaDeviatov2024}. There, instead of the expected for the MnTe antiferromagnet $g-$wave spin-splitting with a characteristic $\frac{2\pi}{3}$ periodicity \cite{MnTe2023exp}, a $\pi$ periodicity ($d-$wave order) of the magnetization has been observed \cite{OrlovaDeviatov2024}. 
We note that the mirror-symmetric spin-splitting appears to be looking like the $d-$wave spin-splitting and can be seen as a $k_{y} = 0$ cut of the $g-$wave spin-splitting $\propto \sigma_{z}k_{z}k_{x}(k_{x}^2 - 3k_{y}^2)$.
In addition, the $d-$wave like symmetric spin-splitting in MnTe may be originating due to either spin-orbit coupling or strain \cite{Belashchenko2024}.
In \cite{Belashchenko2024} it has been claimed that the spin-orbit gives a rather small splitting. 
In \cite{OrlovaDeviatov2024} experiments the samples are not strained.
Thus our theory suggests that mirror-symmetric spin-splitting shown in Fig. (\ref{fig:fig3}) may be relevant to MnTe. 
Hence spin-splitter effect \cite{Gonzalez-Hernandez2021} is expected in MnTe.

The model Eq. (\ref{ModelA}) will also show anomalous spin Hall effect when the N\'{e}el order is in the $z-$direction. The anomalous Hall effect is absent in the system, but passing a spin-polarized electric current through the system will generate transverse to the current voltage drop with spin projection defining the sign of the voltage drop. 
The anomalous Hall effect for spin $\sigma = \uparrow, \downarrow$ is defined by a current of spin-up/down fermions which flows in the transverse direction to the electric field ${\bf E}$, 
\begin{align}\label{currentAHE}
{\bf j}^{\sigma} & = e^2
\left[ \int_{\mathrm{BZ}}\frac{d^2 k}{(2\pi)^2}\sum_{n = \pm} {\bm \Omega}^{\sigma}_{{\bf k};n} 
{\cal F}(\epsilon_{{\bf k};n;s} )\right]\times {\bf E} ,
\end{align}
where $e$ is the electron's charge, $n=\pm$ is the band index denoting conduction (plotted in Fig. (\ref{fig:fig2})) and valence bands correspondingly, ${\bm \Omega}^{\sigma}_{{\bf k};n} = \Omega^{\sigma}_{{\bf k};n}{\bf e}_{z}$ is the Berry curvature of the $\epsilon_{{\bf k};n;\sigma}$ energy band of the Eq. (\ref{ModelA}), 
${\cal F}(\epsilon) = (e^{\frac{\epsilon - \mu}{T}}+1 )^{-1}$ is the Fermi-Dirac distribution function.. We plot $\sigma^{\uparrow} \equiv j_{x
}^{\uparrow}/(e^2E_{y})$ in the right of Fig. (\ref{fig:fig2}) as a function of temperature $T$ for different Fermi levels $\mu$. We checked that indeed $\sigma^{\downarrow} = - \sigma^{\uparrow}$.

\begin{figure}[t] 
\includegraphics[width=0.8 \columnwidth ]{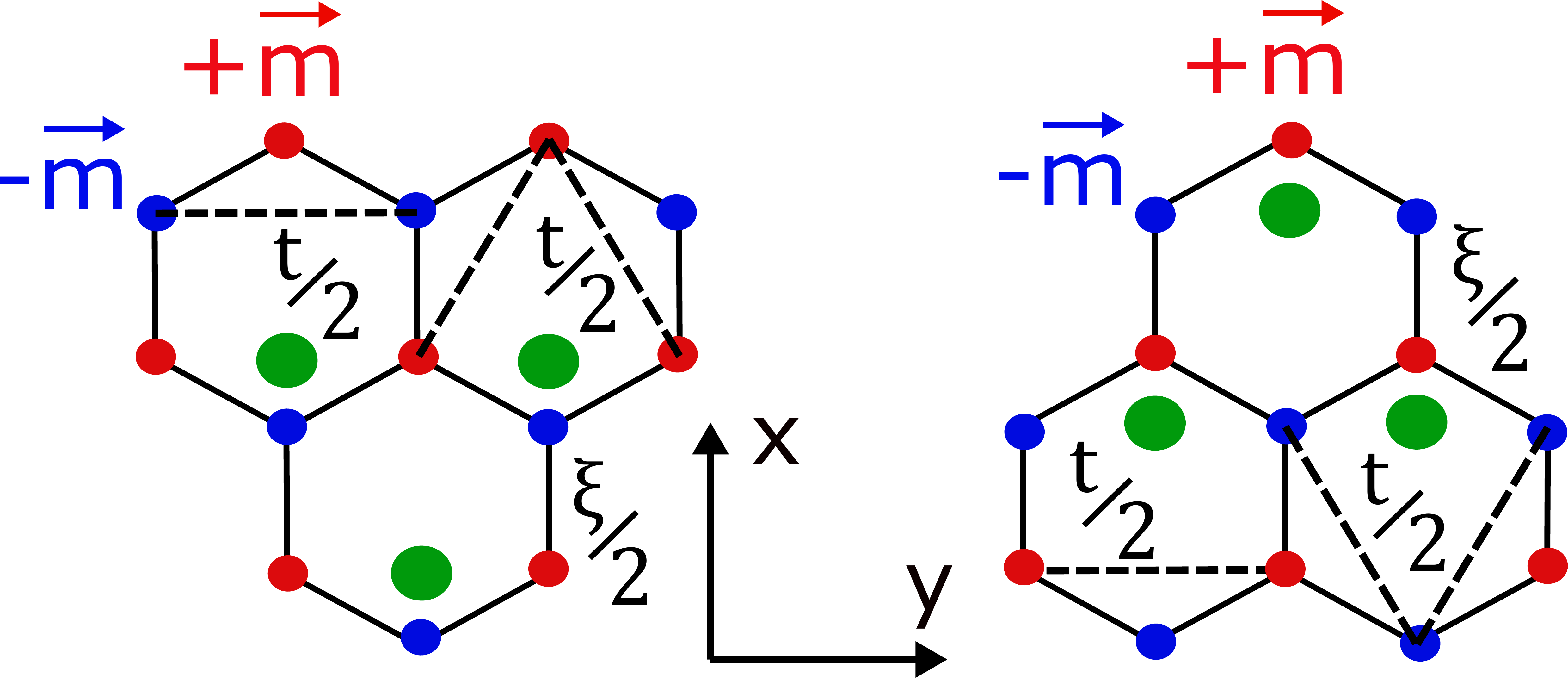} ~~~~

\protect\caption{All the details are the same as in Fig. (\ref{fig:fig1}) with the only difference that the green atom in the unit cell was replaced to a position such that there is no symmetry between the sublattices. The two systems (on the right and on the left) are related by a mirror in $zy$ plane and time-reversal operations. The two systems have a different structure in the second-nearest neighbor hopping.
 }
\label{fig:fig4}  
\end{figure}

It is instructive to study a model in which all symmetries between the sublattices are broken by the position of the green atom. For that let us place the green atom as shown in Fig. (\ref{fig:fig4}). Again, the green atom blocks the tunneling across it and in addition it may result in spin-orbit coupling.
The Hamiltonian of the model is
\begin{align}\label{ModelB}
\hat{H}_{\mathrm{B}}= 
\left[
\begin{array}{cc} 
{\bf m}\cdot{\bm \sigma} + \chi_{\mathrm{R};{\bf k}} \sigma_{z}+ t_{\mathrm{R};{\bf k}}  & \xi_{\bf k} + \eta_{\bf k} \sigma_{z}\\
\xi_{\bf k}^{*} + \eta_{\bf k}^{*} \sigma_{z} & -{\bf m}\cdot{\bm \sigma} + \chi_{\mathrm{B};{\bf k}}\sigma_{z} + t_{\mathrm{B};{\bf k}}
\end{array}
\right],
\end{align}
where $\xi_{\bf k}$ is the same as in the model Eq. (\ref{ModelA}), while second-nearest neighbor hopping structure reads as
\begin{align}
&
t_{\mathrm{R};{\bf k}} = 2t\cos\left( \frac{\sqrt{3}k_{x}}{2} \right)\cos\left( \frac{k_{y}}{2} \right),
\nonumber
\\
&
t_{\mathrm{B};{\bf k}} = t\cos\left( k_{y} \right).
\end{align}

\begin{figure}[t] 
\includegraphics[width=0.4 \columnwidth ]{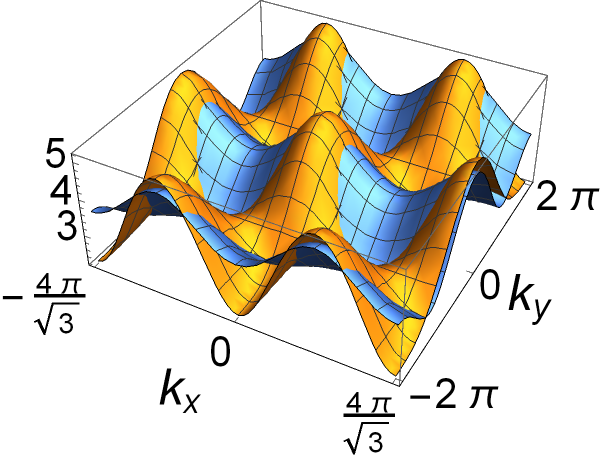} ~~~~
\includegraphics[width=0.4 \columnwidth ]{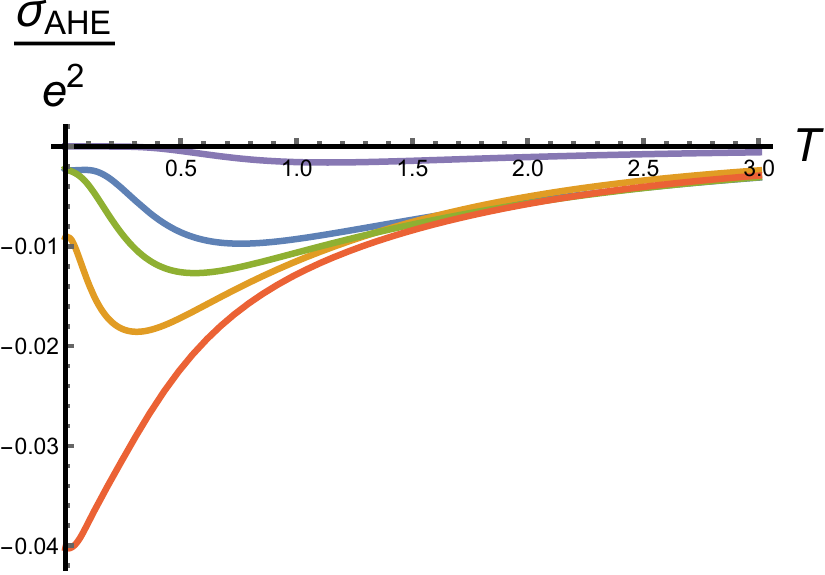}

\protect\caption{Left: spectrum of the conduction band in three times the first Brillouin zone (three pairs of ${\bf K}$ and ${\bf K}^\prime$ points are shown) for illustration purposes. Blue color corresponds to the spin-up fermions, yellow to the spin-down. Parameters are $\xi = 1$, $t=0.5$, $m=3$, $\lambda=\eta = 0$ in units of $\xi$ such that $\xi = 1$.  
Right: plot of the anomalous Hall conductivity for $\xi = 1$, $t=0.5$, $m=3$, $\eta = 0.4$, and $\lambda=0.2$ for 
blue $\mu =4$, green $\mu=3.7$, red $\mu=3$, yellow $\mu=2$, purple $\mu=0$.
 }
\label{fig:fig5}  
\end{figure}

\begin{figure}[t] 

\includegraphics[width=0.3 \columnwidth ]{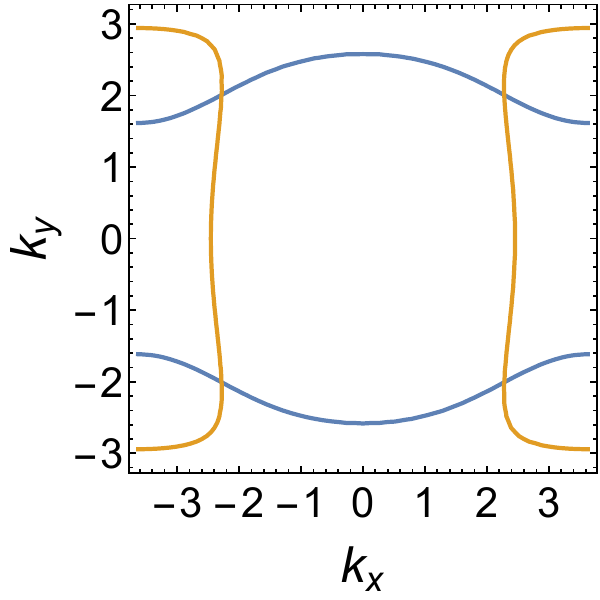} 
\includegraphics[width=0.3 \columnwidth ]{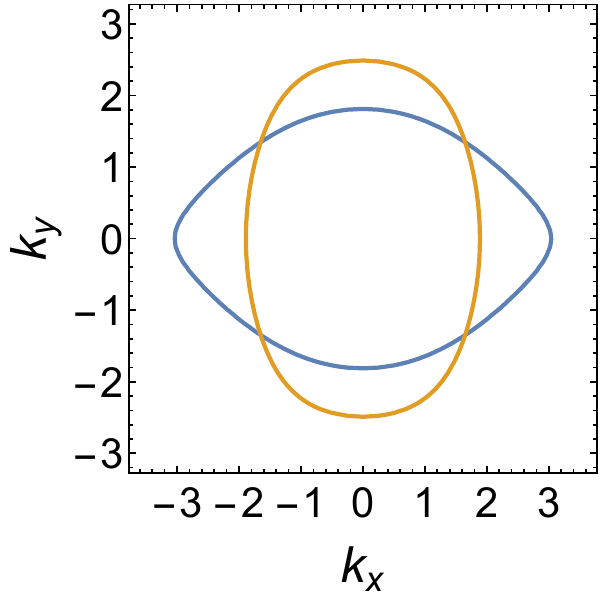} 
\includegraphics[width=0.3 \columnwidth ]{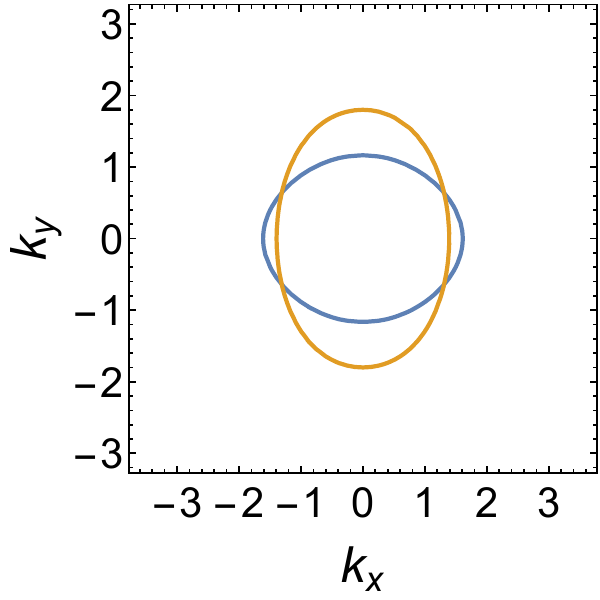} 

\protect\caption{Contour plots of the Fermi surface for chosen Fermi level $\mu$. The two Fermi surfaces are not related to each other by any symmetry operation. Blue color corresponds to the spin-up fermions, yellow to the spin-down.
Both are plotted for  $\xi = 1$, $t=0.5$, $m=3$, $\lambda=\eta = 0$. Left is for $\mu=3$, center is for $\mu=3.7$, and right is for $\mu=4.3$. 
All parameters are in units of $\xi$. 
 }
\label{fig:fig6}  
\end{figure}
The lattice allows for the first-nearest neighbor spin-orbit coupling
$
\eta_{\bf k} = 2\eta e^{i\frac{k_{x}}{2\sqrt{3}}}\sin\left( \frac{k_{y}}{2} \right).
$
The second-nearest neighbor spin-orbit coupling has the same property as in the previous model. 
It is again defined as $
\chi_{\mathrm{R}/\mathrm{B};{\bf k}} = \left( \lambda_{{\bf k}} \pm \gamma_{{\bf k}}\right), 
$
where
\begin{align}
&
\lambda_{\bf k} = \lambda \sin\left( \frac{\sqrt{3}k_{x}}{2} \right)\cos\left( \frac{k_{y}}{2} \right) - \frac{\lambda}{2}  \sin(k_{y}),
\end{align}
and its anisotropic part is
\begin{align}
\gamma_{\bf k} = \lambda \sin\left( \frac{\sqrt{3}k_{x}}{2} \right)\cos\left( \frac{k_{y}}{2} \right) + \frac{\lambda}{2} \sin(k_{y}).
\end{align}
Just like in the previous model $\lambda \approx 0$ if the charges of green and magnetic atoms are of the same sign. In the opposite case, the spin-orbit coupling is enhanced. We set the N\'{e}el order to be in the $z-$direction. The spectrum is plotted in Fig. (\ref{fig:fig5}) and its equal energy contours in 
Fig. (\ref{fig:fig6}). It is clear that spin-up fermions can't be connected with the spin-down by any symmetry operation.

In Ref. \cite{Zyuzin2024} it has been suggested that the charge-density wave order together with the N\'{e}el may result in Zeeman-like spin-splitting of conducting fermions and as a result there may be the anomalous Hall effect in the system upon addition of the spin-orbit coupling. This is the consequence of the broken symmetries between the two sublattices. In the model Eq. (\ref{ModelB}) all symmetries between the two sublattices are broken by the non-magnetic green atom. This allows for the anomalous Hall effect in the system.
The anomalous Hall conductivity is shown to be non-zero only for both $t\neq 0$ and either $\eta \neq 0$ or $\lambda \neq 0$ (or both non-zero). We plot 
$\sigma_{\mathrm{AHE}} \equiv \left( j_{x
}^{\uparrow}+j_{x
}^{\downarrow}\right)/E_{y}$ in the right of Fig. (\ref{fig:fig4}). 
 Just like in Ref. \cite{Zyuzin2024} the magnitude of the conductivity is small. There is no need for the Rashba spin-orbit coupling for the anomalous Hall effect in this system, but there are effects due to its interplay with the spin-splitting \cite{Zyuzin2024} or with the spin-splitting and the Zeeman magnetic field \cite{VorobevZyuzin2024}. We note that certainly there are other contributions to the anomalous Hall effect such as the side-jump and skew-scattering which would alter the magnitude of the effect. Here we just wanted to demonstrate that the effect exists. 
We note that just like in the model Eq. (\ref{ModelA}) the spin-splitting shown in Fig. (\ref{fig:fig6}) is almost $d-$wave symmetric in a sense there are four zeros of the splitting. Materials with the mirror-symmetric spin-splitting shown in Fig (\ref{fig:fig3}) or asymmetric spin-splitting shown in Fig (\ref{fig:fig6}) may be relevant to spintronics as devices with the spin-splitter effect \cite{Gonzalez-Hernandez2021}.

To conclude we have identified a distinct class of antiferromagnets with mirror-symmetric spin-splitting of conducting fermions.
We have also studied its broken mirror symmetry phase.
There is an experimental evidence \cite{OrlovaDeviatov2024} that the studied here model systems may be relevant to the MnTe antiferromagnet.

\textit{Acknowledgements}
The author thanks E.V. Deviatov, N.N. Orlova, and J. Sinova for helpful discussions.
The author is grateful to Pirinem School of Theoretical Physics. This work is supported by FFWR-2024-0016.

\end{document}